# Quaternionic Representations of the Pyritohedral Group, Related Polyhedra and Lattices

**Nazife Ozdes Koca**[a*], **Mehmet Koca**[a], **Aida Al-Mukhaini**[a] and **Amal Al-Qanobi**[a]

[a]Department of Physics, College of Science, Sultan Qaboos University, P.O. Box 36, Al-Khoud, 123 Muscat, Sultanate of Oman. *Correspondence e-mail: nazife@squ.edu.om

We construct the fcc (face centered cubic), bcc (body centered cubic) and sc (simple cubic) lattices as the root and the weight lattices of the affine Coxeter groups $W(D_3)$ and $W(B_3) \approx Aut(D_3)$. The rank-3 Coxeter-Weyl groups describing the point tetrahedral symmetry and the octahedral symmetry of the cubic lattices have been constructed in terms of quaternions. Reflection planes of the Coxeter-Dynkin diagrams are identified with certain planes of the unit cube. It turns out that the pyritohedral symmetry takes a simpler form in terms of quaternionic representation. The $D_3$ diagram is used to construct the vertices of polyhedra relevant to the cubic lattices and, in particular, constructions of the pseudoicosahedron and its dual pyritohedron are explicitly worked out.

## 1. Introduction

The Coxeter-Weyl groups acting as discrete groups in 3D euclidean space generate orbits (Koca et al., 2007; Koca et al., 2010) representing vertices of certain polyhedra which can be used to describe the symmetries of molecular structures (Cotton et al. 1999) and viral symmetries (Caspar and Klug, 1962; Twarock, 2006). Higher dimensional lattices described by the affine extensions of the Coxeter-Weyl groups can be used to describe the symmetries of the quasicrystal structures when projected into lower dimensions (Katz and Duneau, 1986; Elser, 1985; Baake et al., 1990a; Baake et al., 1990b). The rank-3 Coxeter-Weyl groups $W(D_3)$ and $W(B_3) \approx Aut(D_3)$ define the point tetrahedral and octahedral symmetries of the cubic lattices which have enormous number of applications in material science.
 Quaternionic representation of the isometries of the 4D and 3D euclidean spaces, in spite of its attractiveness, has long been neglected. Classification of the finite subgroups of the isometries of the 4D and 3D euclidean spaces in terms of quaternions (Conway and Smith, 2003) stimulates the use of them to describe the 4D polytopes (Koca et al. 2013, Koca et al. 2014a), 3D polyhedra and the related lattices. In this paper we explicitly show that the root lattice and the weight lattice of the affine Coxeter-Weyl group $W_a(D_3)$ describe the fcc and bcc lattices respectively (Conway and Sloane, 1988). We point out that there is a natural correspondence between the octahedral symmetry of the fcc and bcc lattices and the binary octahedral group of quaternions. The simple cubic lattice is described by the affine Coxeter group $W_a(B_3)$. Another concern is the construction of the pyritohedral group and the related polyhedra in terms of quaternions. Construction of the vertices of the pseudoicosahedron with pyritohedral symmetry has been explained by the use of Coxeter-Dynkin diagram $D_3$.

The paper is organized as follows. In Section 2 we introduce the quaternions and their relevance to the $O(4)$ and $O(3)$ transformations. We also give the sets of quaternions defining the binary tetrahedral and binary octahedral groups. We begin with Section 3 by introducing the Coxeter-Dynkin diagrams $A_2$ and $B_2$ by which we construct the hexagonal and square lattices respectively. Extension to the symmetries generated by Coxeter-Dynkin diagrams $D_3$ and $B_3$ in which the simple roots (lattice generating vectors) expressed in terms of imaginary quaternionic units is straightforward. The group generators are written in terms of quaternions and the reflection planes of the diagrams $D_3$ and $B_3$ are identified as certain planes of the unit cube. Section 4 is devoted to the construction of the icosahedron-pseudoicosahedron and its dual polyhedron from the $D_3$ diagram. It is noted that the truncated octahedron, the Wigner-Seitz cell of the bcc lattice, splits into two pseudoicosahedra which are mirror images of each other. Finally in Section 5 we discuss the possible use of pseudoicosahedra and their dual pyritohedra for the descriptions of the crystals possessing pyritohedral symmetry.

## 2. Quaternions and $O(3)$ transformations

A quaternion is a hypercomplex number with three imaginary units $i, j, k$ satisfying the relation $i^2 = j^2 = k^2 = ijk = -1$ (Hamilton, 1853). Hereafter we shall use a different notation for the imaginary units. Let us redefine them as $e_1 = i, e_2 = j, e_3 = k$. Let $q = q_0 + q_i e_i$, ($i = 1,2,3$) be a real unit quaternion with its quaternion conjugate defined by $\bar{q} = q_0 - q_i e_i$ and its norm $q\bar{q} = \bar{q}q = q_0^2 + q_1^2 + q_2^2 + q_3^2 = 1$. The Hamilton's relation can be written in the form
$$e_i e_j = -\delta_{ij} + \varepsilon_{ijk} e_k, \quad (i, j, k = 1,2,3) \tag{1}$$
where $\delta_{ij}$ and $\varepsilon_{ijk}$ are the Kronecker and Levi-Civita symbols and summation over the repeated indices is implicit. Quaternions form a group isomorphic to the unitary group $SU(2)$. With the definition of the scalar product
$$(p, q) = \frac{1}{2}(\bar{p}q + \bar{q}p) = \frac{1}{2}(p\bar{q} + q\bar{p}), \tag{2}$$
quaternions generate the four-dimensional euclidean space.

Let $p$ and $q$ be two arbitrary unit quaternions $p\bar{p} = q\bar{q} = 1$ with six real parameters. The transformations of an arbitrary quaternion $t \to ptq$ and $t \to p\bar{t}q$ define the orthogonal transformation of the group $O(4)$. It is clear that the above transformations preserve the norm $t\bar{t} = t_0^2 + t_1^2 + t_2^2 + t_3^2$. We define the above transformations as abstract group operations by the notations

$$t \to ptq := [p,q]t \ , \ t \to p\bar{t}q := [p,q]^* t \tag{3}$$

dropping also $t$, the pair of quaternions defines a set closed under multiplication.
The inverse elements take the forms (Koca et al, 2001)

$$[p,q]^{-1} = [\bar{p},\bar{q}], \ ([p,q]^*)^{-1} = [\bar{q},\bar{p}]^*. \tag{4}$$

With the choice of $q = \bar{p}$ the orthogonal transformations define a three parameter subgroup $O(3)$. This restriction allows the definition of a quaternion consisting of a scalar component $Sc(t) = t_0$ and a vector component $\text{Vec}(t) = t_1 e_1 + t_2 e_2 + t_3 e_3$. The transformation $[p,\bar{p}]$ and $[p,\bar{p}]^*$ leaves the $Sc(t) = t_0$ invariant. Therefore in 3D Euclidean space one can assume, without loss of generality that the quaternions consist of only vector components that is $t = t_1 e_1 + t_2 e_2 + t_3 e_3$ satisfying $\bar{t} = -t$. With this restriction to the 3D space the element $[p,\bar{p}]^*$ takes a simple form $[p,\bar{p}]^* \Rightarrow [p,-\bar{p}]$.
Therefore the transformations of the group $O(3)$ can be written as $[p,\pm\bar{p}]$ where the group element $[p,\bar{p}]$ represents the rotations around the vector $\text{Vec}(p) = (p_1, p_2, p_3)$ and $[p,-\bar{p}]$ is a rotory inversion (Coxeter, 1973). In the references (Conway and Smith, 2003 and du Val, 1964) one can find the classifications of the finite subgroups of quaternions and the related finite subgroups of the group of orthogonal transformations $O(3)$. We will not list all of them here but simply display the sets of finite subgroups of quaternions related to the tetrahedral and octahedral groups. The set $T$ is given by the group elements

$$T = \{\pm 1, \pm e_1, \pm e_2, \pm e_3, \frac{1}{2}(\pm 1 \pm e_1 \pm e_2 \pm e_3)\} \tag{5}$$

and is called the binary tetrahedral group of order 24. Another set of 24 quaternions defined by

$$T' = \{\frac{1}{\sqrt{2}}(\pm 1 \pm e_1), \frac{1}{\sqrt{2}}(\pm e_2 \pm e_3), \frac{1}{\sqrt{2}}(\pm 1 \pm e_2), \frac{1}{\sqrt{2}}(\pm e_3 \pm e_1), \frac{1}{\sqrt{2}}(\pm 1 \pm e_3), \frac{1}{\sqrt{2}}(\pm e_1 \pm e_2)\} \tag{6}$$

does not form a group by itself since it satisfies the relations $T'T' \subset T$ and $T'T \subset T'$ the set $O = T \cup T'$ forms another finite subgroup of quaternions called the binary octahedral group. We will see that these sets play an essential role in the definition of the tetrahedral and octahedral groups as well as their chiral subgroups and the pyritohedral subgroup.

## 3. The Coxeter-Weyl groups $W(D_3)$ and $W(B_3)$ represented by quaternions

The Coxeter-Weyl groups are generated by reflections with respect to some hyperplanes represented by the vectors (also called roots in the literature of Lie Algebra). The roots $\alpha_1, \alpha_2, ..., \alpha_n$ of the Coxeter-Dynkin diagram are the vectors orthogonal to certain hyperplanes. Denote by $r_{\alpha_i} := r_i$, $(i = 1, 2, ..., n)$ the reflection operator with respect to the hyperplane orthogonal to the root $\alpha_i$. Then the presentation of the Coxeter-Weyl group $W(G)$ is given by

$$W(G) = \langle r_1, r_2, ..., r_n \mid (r_i r_j)^{m_{ij}} = 1 \rangle \tag{7}$$

where $m_{ij}$ is an integer label with $m_{ii} = 1$, $m_{ij} = 2, 3, 4$ and $6$ for $i \neq j$ representing respectively no line, one line, two lines (or label 4) and three lines (or label 6) between the nodes of the Coxeter-Dynkin diagrams which determine the crystallographic point groups in an arbitrary euclidean space. The list of Coxeter-Weyl groups can be found in many fundamental references (Coxeter and Moser, 1965; Bourbaki, 1968; Humphreys, 1990) including the non-crystallographic Coxeter groups which are out of the scope of this work. Before we discuss the details of the groups $W(D_3)$, $W(B_3)$, related polyhedra and the lattices let us study two simple cases which describe certain polygonal symmetries and the related lattices. They are the Coxeter-Dynkin diagrams representing the groups $W(A_2)$ and $W(B_2)$ whose Coxeter-Dynkin diagrams are given in Fig. 1.

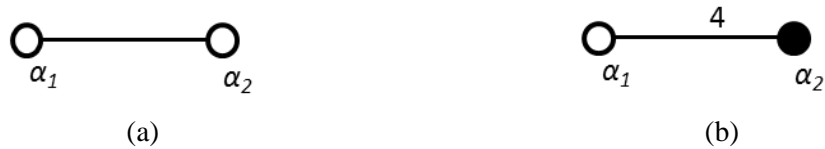

(a)                                    (b)

**Figure 1.** The Coxeter-Dynkin diagrams (a) $A_2$ and (b) $B_2$.

The vectors $\alpha_1$ and $\alpha_2$ (hereafter will be called roots) orthogonal to the planes with respect to which the reflection generators reflect an arbitrary vector $\lambda$ as

$$r_i \lambda = \lambda - \frac{2(\lambda, \alpha_i)}{(\alpha_i, \alpha_i)} \alpha_i, \quad i = 1, 2. \tag{8}$$

The Cartan matrix $C$ (Gram matrix in the lattice terminology) with the matrix elements $C_{ij} = \frac{2(\alpha_i, \alpha_j)}{(\alpha_j, \alpha_j)}$ and the metric $G$ defined by the matrix elements $G_{ij} = (C^{-1})_{ij} \frac{(\alpha_j, \alpha_j)}{2}$ are important for the description of the Coxeter-Weyl groups and the corresponding lattices. The matrices $C$ and $G$ represent the Gram matrices of the direct lattice and the reciprocal lattice respectively (Conway and Sloane, 1988). We take the roots $\alpha_i$ as the generating vectors of the direct lattice. The weights $\omega_i$ spanning the dual space and satisfying the scalar product $(\omega_i, \omega_j) = G_{ij}$ and $(\omega_i, \breve{\alpha}_j) = \delta_{ij}$ with $\breve{\alpha}_j := \dfrac{2\alpha_j}{(\alpha_j, \alpha_j)}$ correspond to the generating vectors of the reciprocal lattice. Now we discuss the lattices associated with each Coxeter-Weyl group.

### 3.1. The lattice determined by the affine group $W_a(A_2)$

The Cartan matrix $C$ and the metric $G$ are given as follows

$$C = \begin{pmatrix} 2 & -1 \\ -1 & 2 \end{pmatrix}, \qquad G = \frac{1}{3}\begin{pmatrix} 2 & 1 \\ 1 & 2 \end{pmatrix}. \tag{9}$$

The plot of the polygons (triangle or hexagon) described by the orbits of the group $W(A_2)$ will be given in orthonormal basis which can be obtained by using the eigenvectors of the Cartan matrix (Koca et al. 2014b)

$$\hat{x}_1 = \frac{1}{\sqrt{2}}(\alpha_1 + \alpha_2), \quad \hat{x}_2 = \frac{1}{\sqrt{6}}(\alpha_1 - \alpha_2), \tag{10}$$

and in the orthonormal basis the roots read as follows:

$$\alpha_1 = (\frac{1}{\sqrt{2}}, \sqrt{\frac{3}{2}}), \quad \alpha_2 = (\frac{1}{\sqrt{2}}, -\sqrt{\frac{3}{2}}). \tag{11}$$

The group generators act on the roots as, $r_1\alpha_1 = -\alpha_1$, $r_1\alpha_2 = \alpha_2 + \alpha_1$, $r_2\alpha_1 = \alpha_1 + \alpha_2$, $r_2\alpha_2 = -\alpha_2$ which generate the group $W(A_2)$ of order 6 isomorphic to the dihedral group $D_3 \approx S_3$. (It is unfortunate that the same letter $D_n$ is used both for the dihedral group and the Coxeter-Dynkin diagram for the $D$- series). Including the Dynkin diagram symmetry which implies $\gamma : \alpha_1 \leftrightarrow \alpha_2$, the Coxeter group can be extended to the dihedral group $D_6$ of order 12. The matrix representations of the generators in the root space $\alpha_1$ and $\alpha_2$ can be written as

$$r_1 = \begin{pmatrix} -1 & 0 \\ 1 & 1 \end{pmatrix}, \quad r_2 = \begin{pmatrix} 1 & 1 \\ 0 & -1 \end{pmatrix}, \quad \gamma = \begin{pmatrix} 0 & 1 \\ 1 & 0 \end{pmatrix}. \tag{12}$$

The affine Coxeter group $W_a(A_2)$ includes another generator $r_0$ which reflects the vectors with respect to a plane bisecting the vector $\alpha_1 + \alpha_2$. This corresponds to a reflection with respect to the root $\alpha_1 + \alpha_2$ followed by a translation by $(\alpha_1 + \alpha_2)$. We shall elaborate this point in the next section in more details. The repeated applications of the generators will generate the root lattice $A_2$ where an arbitrary lattice vector is given by $p = b_1\alpha_1 + b_2\alpha_2$, $b_1, b_2 \in \mathbf{Z}$. The root system consisting of the vectors $\pm\alpha_1, \pm\alpha_2, (\pm\alpha_1 \pm \alpha_2)$ determine the unit cell of the direct lattice which is a hexagon. The lattice is the infinite set of hexagons as shown in Fig. 2. It is interesting to note that the honeycomb lattice exists as graphene made of carbon atoms (Geim, M & Novosolev, S., 2007).

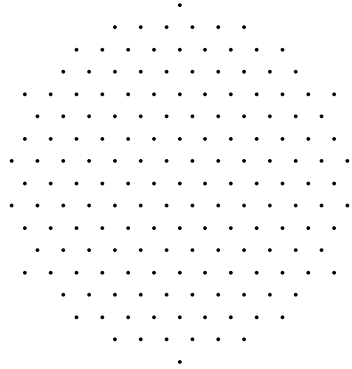

**Figure 2.** The honeycomb lattice.

The weight vectors can be determined from the relation $\omega_i = \sum_{j=1}^{2} G_{ij} \alpha_j$ where the reciprocal lattice $A_2^*$ (weight lattice) vectors are given by $q = a_1\omega_1 + a_2\omega_2 := (a_1 a_2)$. No comma is used between $a_1, a_2 \in \mathbf{Z}$ as long as they do not exceed 10. We will define an orbit in the dual space (reciprocal space) as $W(A_2)(a_1 a_2) := (a_1 a_2)_{A_2}$ which represents an isogonal hexagon with two edge lengths in general. When we have $a_1 = a_2 = 1$ the orbit is a regular hexagon $(11)_{A_2}$ determined by the root system which is invariant under the larger group, the group $Aut(A_2) \approx W(A_2):C_2$. Here (:) represents the semidirect product of two groups where the group on the left is the invariant subgroup of the group $Aut(A_2)$ generated by the matrices in (12). The orbits $(a_1 0)_{A_2}$ and $(0\, a_2)_{A_2}$ represent equilateral triangles of edge lengths $a_1$ and $a_2$ respectively. The union of the orbits $(10)_{A_2} \cup (01)_{A_2}$, each representing an equilateral triangle, constitutes another hexagon dual to the hexagon $(11)_{A_2}$ and describes the unit cell of the weight lattice. Since both the root lattice and the weight lattice are made of hexagons they can be transformed to each other by a change of scale and rotation. Note that the hexagon described by two triangles is invariant under the group $Aut(A_2)$ which involves the Dynkin diagram symmetry.

We assume that the reader is familiar with the concept of nearest neighbor region which is called the Wigner-Seitz cell or the first Brillouin zone by the crystallographers but also known as the Voronoi cell or Drichlet region by the mathematicians. The Wigner-Seitz cell of the root lattice $A_2$ is the cell $(10)_{A_2} \cup (01)_{A_2}$ which can be determined from its primitive cell as shown in Fig. 3. The Wigner-Seitz cell $(10)_{A_2} \cup (01)_{A_2}$ is a scaled copy of the dual of the primitive cell $(11)_{A_2}$ after a $30^0$ rotation. It is obvious that the unit cell of the weight lattice is the hexagon $(10)_{A_2} \cup (01)_{A_2}$. It can be shown that the Wigner-Seitz cell of the weight lattice is the orbit $\frac{1}{3}(11)_{A_2}$ which is a scaled copy of the root polyhedron (hexagon in this case) by the Coxeter number 3 of the group $W(A_2)$. It can be shown that this is a general property of the $A_n$ series of the Coxeter-Weyl groups. Here we should emphasize that the root lattice and the weight lattice of $A_2$ are identical up to a scale transformation and a group action.

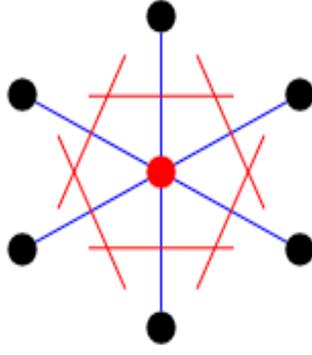

**Figure 3.** The Wigner-Seitz cell $(10)_{A_2} \cup (01)_{A_2}$ inscribed in the primitive cell of the root lattice.

### 3.2. The lattice determined by the affine group $W_a(B_2)$

The roots of $B_2$ consist of long and short roots; in a particular orthonormal system with $(l_i, l_j) = \delta_{ij}$ the simple roots can be written as $\alpha_1 = l_1 - l_2, \alpha_2 = l_2$ [Koca et al 2015]. The Cartan matrix, its inverse and the matrix $G$ are given by

$$C = \begin{pmatrix} 2 & -2 \\ -1 & 2 \end{pmatrix}, \quad C^{-1} = \begin{pmatrix} 1 & 1 \\ \frac{1}{2} & 1 \end{pmatrix}, \quad G = \begin{pmatrix} 1 & \frac{1}{2} \\ \frac{1}{2} & \frac{1}{2} \end{pmatrix}. \tag{13}$$

The group generators $r_1$ and $r_2$ act in the root space as

$$r_1 \alpha_1 = -\alpha_1, \; r_1 \alpha_2 = \alpha_1 + \alpha_2, \; r_2 \alpha_1 = \alpha_1 + 2\alpha_2, \; r_2 \alpha_2 = -\alpha_2 \tag{14}$$

which can be represented by the matrices

$$r_1 = \begin{pmatrix} -1 & 0 \\ 1 & 1 \end{pmatrix}, \quad r_2 = \begin{pmatrix} 1 & 2 \\ 0 & -1 \end{pmatrix}. \tag{15}$$

They generate the dihedral group $D_4$ of order 8 which is the symmetry of the square. The root system consists of two orbits $\{\pm\alpha_2, \pm(\alpha_1+\alpha_2)\}$ and $\{\pm\alpha_1, \pm(\alpha_1+2\alpha_2)\}$ which can also be written in the orthonormal basis as $(10)_{B_2} = \{\pm l_1, \pm l_2\}$ and $(02)_{B_2} = \{\pm l_1 \pm l_2\}$ corresponding to two squares respectively. It is clear that the root lattice is generated by the short roots of $B_2$. The unit cell is the square represented by the vectors $\{\pm l_1, \pm l_2\}$. The affine generator is obtained as the reflection $r_0$ followed by a translation of any vector by the highest short root $\alpha_1+\alpha_2 = l_1$. Therefore the root lattice is represented either by the vectors $b_1\alpha_1+b_2\alpha_2$; $b_1,b_2 \in \mathbf{Z}$ or equivalently by $m_1 l_1 + m_2 l_2$; $m_1, m_2 \in \mathbf{Z}$. The Wigner–Seitz cell is a square with the vertices $(01)_{B_2} = \{\pm\frac{1}{2}l_1 \pm \frac{1}{2}l_2\}$. The square lattice generated by the short roots of $B_2$ is shown in Fig. 4.

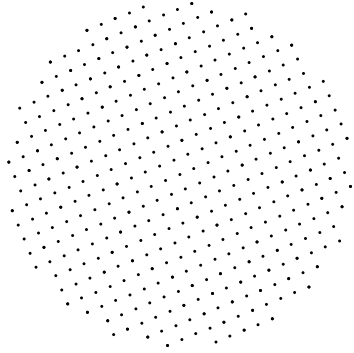

**Figure 4.** The square lattice as the root lattice of $B_2$.

### 3.3. Construction of the fcc and the bcc lattices as the affine Coxeter-Weyl groups $W_a(D_3)$

The Coxeter-Dynkin diagram of $D_3$ with the quaternionic simple roots is given by Fig. 5.

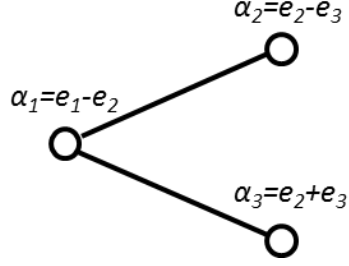

**Figure 5.** The Coxeter-Dynkin diagram $D_3$ with quaternionic simple roots. The angle between two connected roots is $120^0$ otherwise they are orthogonal.

Note that $D_3$ and $A_3$ describes the same Coxeter-Dynkin diagram. An arbitrary quaternion $\lambda$ when reflected by the operator $r_\alpha$ with respect to the hyperplane orthogonal to the quaternion $\alpha$ the formula (8) is given in terms of quaternion multiplication (Koca et al. 2001) as

$$r_\alpha \lambda = -\frac{\alpha}{\sqrt{2}} \bar{\lambda} \frac{\alpha}{\sqrt{2}} := [\frac{\alpha}{\sqrt{2}}, -\frac{\alpha}{\sqrt{2}}]^* \lambda . \tag{16}$$

We used the definition of the scalar product of (2).

The Cartan matrix of the Coxeter-Dynkin diagram $D_3$ and its inverse matrix are given respectively by the matrices

$$C = \begin{bmatrix} 2 & -1 & -1 \\ -1 & 2 & 0 \\ -1 & 0 & 2 \end{bmatrix}, \quad C^{-1} = \frac{1}{4}\begin{bmatrix} 4 & 2 & 2 \\ 2 & 3 & 1 \\ 2 & 1 & 3 \end{bmatrix} . \tag{17}$$

The generators of the Coxeter group $W(D_3)$ are then given in the notation of (16) by

$$\begin{aligned}
r_1 &= [\frac{1}{\sqrt{2}}(e_1 - e_2), -\frac{1}{\sqrt{2}}(e_1 - e_2)]^* = [\frac{1}{\sqrt{2}}(e_1 - e_2), \frac{1}{\sqrt{2}}(e_1 - e_2)], \\
r_2 &= [\frac{1}{\sqrt{2}}(e_2 - e_3), -\frac{1}{\sqrt{2}}(e_2 - e_3)]^* = [\frac{1}{\sqrt{2}}(e_2 - e_3), \frac{1}{\sqrt{2}}(e_2 - e_3)], \\
r_3 &= [\frac{1}{\sqrt{2}}(e_2 + e_3), -\frac{1}{\sqrt{2}}(e_2 + e_3)]^* = [\frac{1}{\sqrt{2}}(e_2 + e_3), \frac{1}{\sqrt{2}}(e_2 + e_3)].
\end{aligned} \tag{18}$$

It is straightforward to check that the reflection generators transform the quaternionic imaginary units as follows:

$$r_1: e_1 \leftrightarrow e_2, e_3 \rightarrow e_3; \quad r_2: e_1 \rightarrow e_1, e_2 \leftrightarrow e_3; \quad r_3: e_1 \rightarrow e_1, e_2 \leftrightarrow -e_3. \tag{19}$$

They generate the Coxeter-Weyl group $W(D_3)$ of order 24 isomorphic to the tetrahedral group, the elements of which can be written compactly by the set

$$W(A_3) = \{[p, \bar{p}] \cup [t, -\bar{t}]\}, \quad p \in T, \ t \in T'. \tag{20}$$

or more compactly by the notation $W(A_3) = \{[T, \bar{T}] \cup [T', -\bar{T}']\}$. This is a hybrid group also called the **tetra-octahedral** group (Conway and Smith, 2003). Here $T$ and $T'$ are the sets of quaternions given in (5-6).

When the simple roots are chosen as $\alpha_1 = e_1 - e_2, \alpha_2 = e_2 - e_3, \alpha_3 = e_2 + e_3$ as in Fig. 5 then the weight vectors are determined as

$$\omega_1 \equiv (100) = e_1, \quad \omega_2 \equiv (010) = \frac{1}{2}(e_1 + e_2 - e_3), \quad \omega_3 \equiv (001) = \frac{1}{2}(e_1 + e_2 + e_3). \tag{21}$$

Using the orbit definition $W(D_3)(a_1 a_2 a_3) := (a_1 a_2 a_3)_{D_3}$ some of the important orbits are given as follows

$$\begin{aligned}
(100)_{D_3} &= \{\pm e_1, \pm e_2, \pm e_3\}, \\
(010)_{D_3} &= \{\frac{1}{2}(-e_1 - e_2 - e_3), \frac{1}{2}(-e_1 + e_2 + e_3), \frac{1}{2}(e_1 + e_2 - e_3), \frac{1}{2}(e_1 - e_2 + e_3)\}, \\
(001)_{D_3} &= \{\frac{1}{2}(e_1 + e_2 + e_3), \frac{1}{2}(e_1 - e_2 - e_3), \frac{1}{2}(-e_1 - e_2 + e_3), \frac{1}{2}(-e_1 + e_2 - e_3)\}, \\
(011)_{D_3} &= \{\pm e_1 \pm e_2, \pm e_2 \pm e_3, \pm e_3 \pm e_1\}.
\end{aligned} \tag{22}$$

They respectively represent the vertices of an octahedron, a tetrahedron and another tetrahedron dual to the first tetrahedron. The last orbit $(011)_{D_3}$ represents the root system of the $A_3 \approx D_3$ Lie algebra and also constitutes the vertices of a cuboctahedron [Koca et al, 2007]. The union of the orbits $(010)_{D_3}$ and $(001)_{D_3}$ represents the vertices of a cube. Therefore the symmetry of the union of the orbits $(010)_{D_3} \cup (001)_{D_3}$ requires the Dynkin diagram symmetry $\gamma: \omega_1 \rightarrow \omega_1, \omega_2 \leftrightarrow \omega_3$, or equivalently, $\gamma: \alpha_1 \rightarrow \alpha_1, \alpha_2 \leftrightarrow \alpha_3$, which leads to the transformation on the imaginary quaternions $\gamma: e_1 \rightarrow e_1, e_2 \rightarrow e_2, e_3 \rightarrow -e_3$. In our formulation the Dynkin diagram symmetry operator reads $\gamma = [e_3, -e_3]^* = [e_3, e_3]$ which extends the Coxeter group $W(D_3)$ to the octahedral group $Aut(D_3) \approx W(D_3):C_2$, the automorphism group of the root system of $D_3$. The generators in (18) and $\gamma = [e_3, e_3]$ will lead to the quaternionic representation of the octahedral group of order 48

$$O_h \approx Aut(D_3) = \{[T, \pm\bar{T}] \cup [T', \pm\bar{T}']\}. \tag{23}$$

Note that from now on we are using the symbols of the sets representing their elements.
It is obvious that the maximal subgroups of the octahedral group $Aut(D_3) \approx O_h$ are the groups, each of which of order 24:

Chiral octahedral group : $O \approx \{[T,\bar{T}] \cup [T',\bar{T}']\}$,
Tetrahedral group       : $T_d \approx W(D_3) = \{[T,\bar{T}] \cup [T',-\bar{T}']\}$, (24)
Pyritohedral group      : $T_h = \{[T,\bar{T}] \cup [T,-\bar{T}]\}$.

Below we discuss the geometry of the generators of the group $Aut(D_3) \approx O_h$. Let us denote by the letters $(ABCD)$ and $(A'B'C'D')$ the vertices of the cube shown in Fig. 6a. Each set of four letters represents a tetrahedron inscribed in the cube.

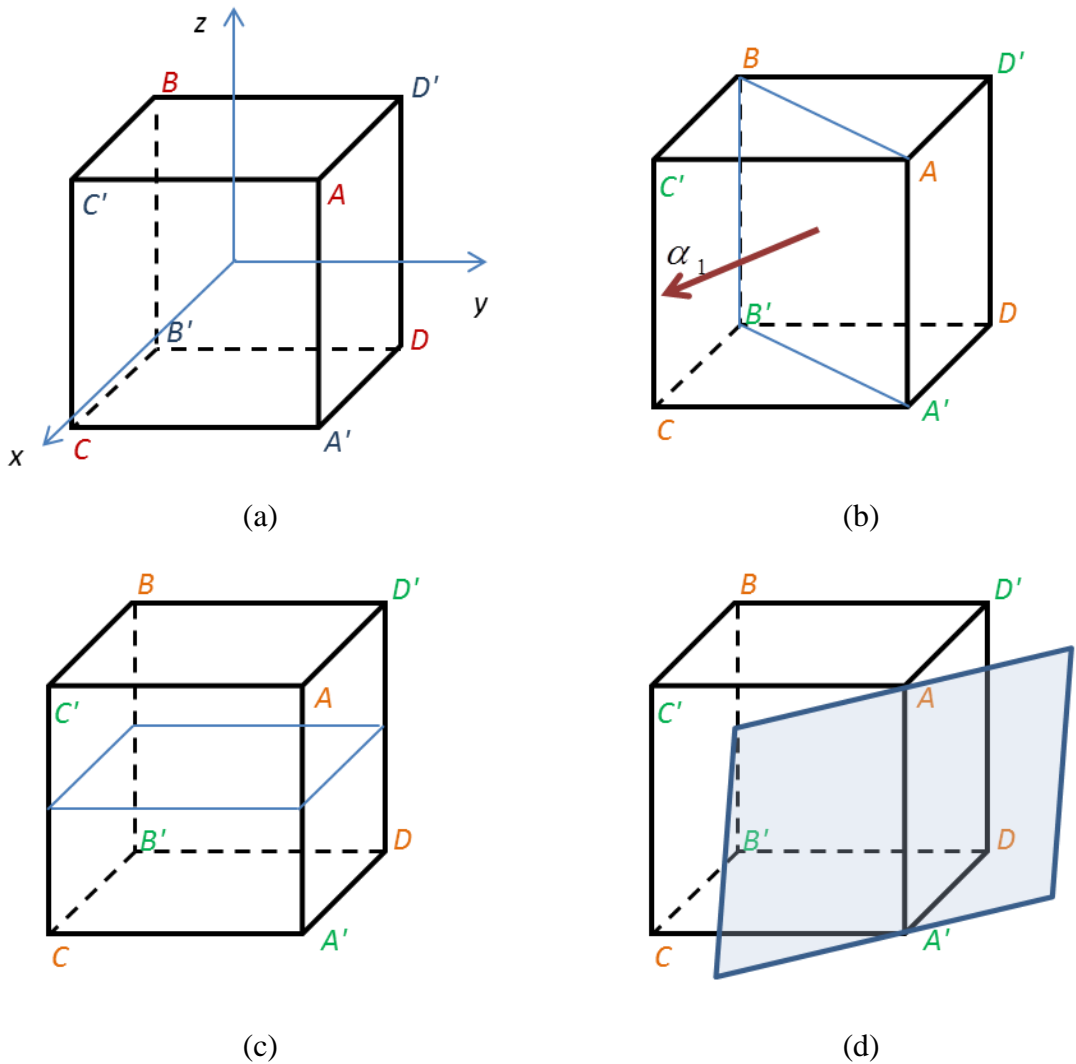

(a)      (b)

(c)      (d)

**Figure 6.** (a) The unit cube with the origin of the coordinate system (*xyz*) is at the center, (b) The plane orthogonal to the simple root $\alpha_1$, (c) The reflection plane represented by the Dynkin diagram symmetry $\gamma$, (d) The affine reflection plane represented by the generator $r_0$.

The letters represent the quaternionic vertices as follows:

$$A = \frac{1}{2}(e_1+e_2+e_3), B = \frac{1}{2}(-e_1-e_2+e_3), C = \frac{1}{2}(e_1-e_2-e_3), D = \frac{1}{2}(-e_1+e_2-e_3)$$
$$A' = \frac{1}{2}(e_1+e_2-e_3), B' = \frac{1}{2}(-e_1-e_2-e_3), C' = \frac{1}{2}(e_1-e_2+e_3), D' = \frac{1}{2}(-e_1+e_2+e_3). \quad (25)$$

It is clear now that the reflection generators of the Coxeter-Weyl group $W(D_3)$ can be written in the permutation notations

$$r_1 = \begin{bmatrix} A & B & C & D \\ A & B & D & C \end{bmatrix}, r_2 = \begin{bmatrix} A & B & C & D \\ A & D & C & B \end{bmatrix}, r_3 = \begin{bmatrix} A & B & C & D \\ C & B & A & D \end{bmatrix}. \quad (26)$$

The same generators can be represented by replacing the letters $(ABCD)$ by their corresponding letters $(A'B'C'D')$, along with a change $r_2 \leftrightarrow r_3$. These generators represent reflections with respect to the planes orthogonal to their corresponding simple roots $\alpha_1, \alpha_2$ and $\alpha_3$. They are respectively the planes $(ABB'A'), (ACB'D')$ and $(BC'A'D)$ as shown in Fig. 6b. The generators of (26) are called transpositions in the terminology of permutation group (Armstrong, 1988) and can also be represented by the transpositions $r_1 = (CD), r_2 = (BD)$ and $r_3 = (AC)$. Two particular generators are sufficient to generate the group of permutations. One can take one transposition, say $r_1 = (CD)$ and a cyclic transformation of all letters (*n-cycle*) which we can choose as $r_1r_2r_3 = (ADBC)$. It represents the permutation in the order $A \to D \to B \to C \to A$. In this notation $r_1$ and $r_1r_2r_3$ represent the generators of the group $S_4$ permuting the four letters. This proves that the tetrahedral group can be represented by the isomorphic groups $T_d \approx W(D_3) \approx S_4 \approx A_4 : C_2$. The latter notation implies that the permutation group is the semi-direct product of the group of even permutations $A_4$ with its normalizer group $C_2$. Now we would like to discuss the role of the Dynkin diagram symmetry generator $\gamma$. It is clear from its definition that it represents a reflection with respect to the *x-y* plane (see Fig.6c) transforming two tetrahedra to each other as $A \leftrightarrow A', B \leftrightarrow B', C \leftrightarrow C'$ and $D \leftrightarrow D'$. When we discuss the Coxeter-Weyl group $W(B_3) \approx O_h$ we will see that the generator $\gamma$ can be represented as one of the generator of the group $W(B_3) \approx O_h$. Next, we would like to discuss the affine extension of the group $W(D_3)$.

Let us introduce the following quantities. For each root $\alpha$ and each integer $k$, we define an affine hyperplane (Humphreys, 1990)

$$H_{\alpha,k} := \{\lambda \in V \,|\, (\lambda, \alpha) = k\} \qquad (27)$$

where $\lambda$ is an arbitrary vector in the euclidean space $V$. Note that $H_{\alpha,k} = H_{-\alpha,-k}$ and $H_{\alpha,0} = H_\alpha$ coincides with the hyperplane passing through the origin and orthogonal to the root $\alpha$. One can define the corresponding affine reflection as

$$r_{\alpha,k}(\lambda) := \lambda - \frac{2((\lambda, \alpha) - k)}{(\alpha, \alpha)} \alpha. \qquad (28)$$

This reflection fixes $H_{\alpha,k}$ pointwise and interchanges the vectors $0 \leftrightarrow \frac{2k\alpha}{(\alpha, \alpha)}$. Note also that the affine reflection can also be written as $t(\frac{2k\alpha}{(\alpha,\alpha)})r_\alpha$ where $t(\lambda)$ sends an arbitrary vector $\mu \to \mu + \lambda$. The affine Coxeter group $W_a$ is generated by all affine reflections $r_{\alpha,k}$ where $\alpha$ is an arbitrary root and $k \in \mathbf{Z}$. It is possible to generate the group with a minimum number of generators. In addition to the generators $r_i$ $(i=1,2,...,n)$, we introduce one more generator, usually denoted by $r_{\alpha_0} := r_0$, where $-\alpha_0 = \tilde{\alpha}$ is the highest root. The extended Coxeter-Dynkin diagram of $D_3$ is illustrated in Fig. 7. Here $\alpha_0$ represents the root of the extended Coxeter-Dynkin diagram. In the following we determine the affine group $W_a(D_3)$ and the associated lattices.

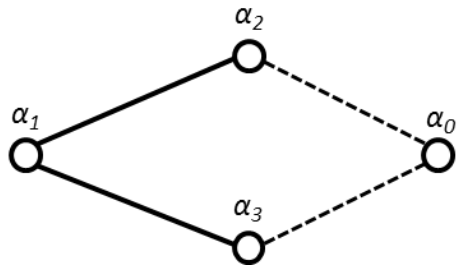

**Figure 7.** Extended Dynkin diagram of $D_3$.

The highest root of $D_3$ can be given in terms of quaternions as $\tilde{\alpha} = (\alpha_1 + \alpha_2 + \alpha_3) = e_1 + e_2$. We define the affine plane $H_{\tilde{\alpha},1} := \{\lambda \in V \,|\, (\lambda, \tilde{\alpha}) = 1\}$ and the reflection generator $r_0 = r_{\tilde{\alpha},1}(\lambda) := \lambda - \frac{2((\lambda,\tilde{\alpha})-1)}{(\tilde{\alpha},\tilde{\alpha})}\tilde{\alpha}$. This generator represents the reflection with respect to the plane bisecting the highest root $\tilde{\alpha} = e_1 + e_2$. In Fig. 6d it is the plane parallel to the plane $CDD'C'$ but shifted in the direction up to $\frac{e_1+e_2}{2}$, that is, tangent to the cube at the edge $AA'$. Reflection with respect to this plane can also be represented by $t(\tilde{\alpha})r_{\tilde{\alpha}}$, corresponding to the reflection with respect to the plane $CDD'C'$ followed by a translation by the vector $\tilde{\alpha} = e_1 + e_2$. Since $r_0$ involves a translation the affine Coxeter group $W_a(D_3)$ is generated by the four generators $\langle r_0, r_1, r_2, r_3 \rangle$. With the inclusion of the Dynkin diagram symmetry the affine group will define a lattice with the octahedral point symmetry.

If we take the generating vectors as the roots $\alpha_1, \alpha_2$ and $\alpha_3$ the affine group generate the fcc lattice (the root lattice $D_3$) where a general lattice vector is given by $p = b_1\alpha_1 + b_2\alpha_2 + b_3\alpha_3$, $b_i \in \mathbf{Z}$. A vector in the root lattice can also be written in terms of the quaternionic imaginary units as $p = m_1 e_1 + m_2 e_2 + m_3 e_3$ where $\sum_{i=1}^{3} m_i = 2b_3 =$ even integer. This is the standard definition of the fcc lattice, namely, the sum of the integers $(m_1, m_2, m_3)$ in the orthonormal bases is equal to even number. The orbit $(011)_{D_3}$, the set of root system of $D_3$, representing the centers of the edges of a cube has a fundamental meaning. It is the nonconventional unit cell of the fcc lattice (Ashcroft & Mermin, 1976). Moreover its dual polyhedron is the union of the orbits $(100)_{D_3} \cup \{(010)_{D_3} \cup (001)_{D_3}\}$ (Koca et al, 2010) which represents a rhombic dodecahedron as shown in Fig. 8. So the dual polyhedron of the non conventional cell of the fcc lattice is the Wigner-Seitz cell.

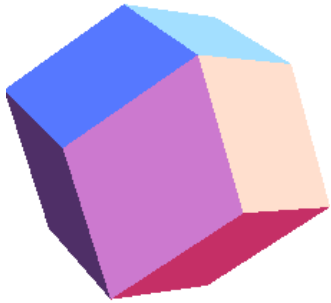

**Figure 8.** Rhombic dodecahedron, the Wigner –Seitz cell of the fcc lattice.

With the choice of the weight vectors in (21) the affine octahedral group generates the weight lattice $D_3^*$ which is the reciprocal lattice of the root lattice, that is the bcc lattice. A general vector of the bcc lattice is given by $q = a_1\omega_1 + a_2\omega_2 + a_3\omega_3$, $a_i \in \mathbf{Z}$. In terms of quaternions it can be written as $q = n_1 e_1 + n_2 e_2 + n_3 e_3$, with either all $n_i$ integers or half integers. Choosing the generating vector twice the weight vectors then an arbitrary bcc vector has the same form as above with $n_i$ are either all even integers or all odd integers. This is the usual definition of the bcc lattice in the literature. The unit cell, a cube including its center, can be represented by the union of the orbits $(010)_{D_3} \cup (001)_{D_3} \cup (000)_{D_3}$. The last orbit is the origin of the coordinate system representing the center of the cube of unit length. The Wigner-Seitz cell of the bcc lattice is the truncated octahedron which is represented by the orbit $\frac{1}{4}(111)_{D_3}$ consisting of 24 vertices given by

$$\frac{1}{4}(111)_{D_3} = \frac{1}{4}\{\pm 2e_1 \pm e_2, \pm 2e_2 \pm e_3, \pm 2e_3 \pm e_1, \pm e_1 \pm 2e_2, \pm e_2 \pm 2e_3, \pm e_3 \pm 2e_1\}. \tag{29}$$

The factor 4 in the denominator is the Coxeter number of the group $W(D_3)$. The truncated octahedron constructed from the vertices (29) is depicted in Fig. 9.

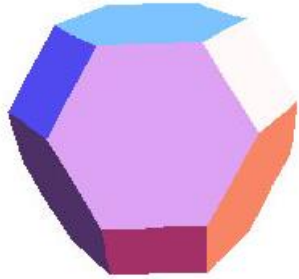

**Figure 9.** Truncated octahedron, the Wigner-Seitz cell of the bcc lattice.

### 3.4. Construction of the sc lattice with its affine Coxeter-Weyl group $W_a(B_3)$

The Coxeter-Dynkin diagram $B_3$ is illustrated in Fig. 10. The simple roots can be taken as $\alpha_1 = e_1 - e_2$, $\alpha_2 = e_2 - e_3$, $\alpha_3 = e_3$. Note that the last root is a short root and the angle between $\alpha_3$ and $\alpha_2$ is $135^0$.

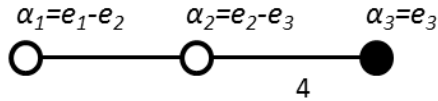

**Figure 10**. The Coxeter-Dynkin diagram $B_3$.

Here the Coxeter-Weyl group $W(B_3)$ is generated by the generators

$$r_1 = [\frac{1}{\sqrt{2}}(e_1 - e_2), \frac{1}{\sqrt{2}}(e_1 - e_2)],\ r_2 = [\frac{1}{\sqrt{2}}(e_2 - e_3), \frac{1}{\sqrt{2}}(e_2 - e_3)],\ r_3 = [e_3, e_3]. \tag{30}$$

They generate the octahedral group $W(B_3) \approx Aut(D_3) = \{[T, \pm\bar{T}] \cup [T', \pm\bar{T}']\}$. It is interesting to note that the first two generators in (30) are identical to the first two generators of the group $W(D_3)$ in (18) and the third generator is identical to the Dynkin diagram symmetry generator $\gamma$. This indicates that the third generator of the group $W(D_3)$ is redundant in the construction of $Aut(D_3)$. One can easily check that the third generator of the tetrahedral group $W(D_3)$ can be written as $\gamma r_2 \gamma$. The weight vectors of $B_3$ are given by

$$\omega_1 = e_1, \ \omega_2 = e_1 + e_2, \ \omega_3 = \frac{1}{2}(e_1 + e_2 + e_3). \tag{31}$$

The orbits $(100)_{B_3}, (010)_{B_3}$ and $(001)_{B_3}$ respectively represent an octahedron, a cuboctahedron and a cube respectively. The truncated octahedron is represented by the orbit $(110)_{B_3}$ which is equivalent to the orbit $(111)_{D_3}$ obtained as the orbit of the group $W(D_3)$. Note that the diagram $B_3$ consists of long roots and short roots of norms $\sqrt{2}$ and 1 respectively. The generators of the octahedral group $W(B_3) \approx O_h$ in (30) generate the root system consisting of the roots

$$(100)_{B_3} \cup (010)_{B_3} = \{\pm e_1, \pm e_2, \pm e_3\} \cup \{\pm e_1 \pm e_2, \pm e_2 \pm e_3, \pm e_3 \pm e_1\}. \tag{32}$$

The short roots represent the centers of the faces (vertices of an octahedron) and the centers of the edges of a cube of 2 unit length. Before we proceed further we would like to discuss the structure of the octahedral group $W(B_3) \approx O_h$ in a little more detail. The proper subgroup (the chiral octahedral group) of the group $W(B_3) \approx O_h$ is generated by the rotation generators $a = r_1 r_2$, $b = r_1 r_3$ satisfying the generation relation $a^3 = b^4 = (ab)^2 = 1$. This relation is sufficient to prove that the chiral octahedral group can be generated by the generators $a$ and $b$. Its geometry can be described with the use of unit cube in Figure 6. The generators $a$ and $b$ permutes the imaginary quaternions in the manner $a: (e_1 e_2 e_3)$ and $b: e_1 \to e_1, e_2 \leftrightarrow -e_3$. They permute the diagonals of the cube as follows:

$$\begin{aligned} a: AB' \to AB', \ CD' \to DC' \to BA' \to CD' \\ b: AB' \to DC' \to CD' \to BA' \to AB'. \end{aligned} \tag{33}$$

This proves that the generators $a$ and $b$ represent the 3-cycle and 4-cycle respectively on four diagonals and generate the symmetric group $S_4$, in other words, $a$ and $b$ generate the chiral octahedral group $O \approx \{[T, \bar{T}] \cup [T', \bar{T}']\} \approx S_4$. The generator $a$ represents a rotation by $120^0$ around the diagonal $AB'$. In terms of quaternion representation it reads $a = [\frac{1}{2}(1+e_1+e_2+e_3), \frac{1}{2}(1-e_1-e_2-e_3)]$. Similarly the generator $b$ represents a rotation in the counterclockwise direction by $90^0$ around $x$ - axis and given in terms of quaternions as $b = [\frac{1}{\sqrt{2}}(1+e_1), \frac{1}{\sqrt{2}}(1-e_1)]$. Moreover the group element $ab = [\frac{1}{\sqrt{2}}(e_1+e_2), -\frac{1}{\sqrt{2}}(e_1+e_2)]$ represents a rotation $180^0$ around the vector $e_1+e_2$. Another generator which commutes with $a$ and $b$ is the group element $(r_1 r_2 r_3)^3$ which inverts the vectors $e_1 \rightarrow -e_1, e_2 \rightarrow -e_2, e_3 \rightarrow -e_3$. It is represented by the group element $[1,-1]$ in the quaternionic notation. Therefore the octahedral group $W(B_3) = \{[T, \pm \bar{T}] \cup [T', \pm \bar{T}']\} \approx S_4 \times C_2$ is the direct product of the permutation group of 4 diagonals and the inversion symmetry $C_2$ generated by $[1,-1]$ which reverses the diagonals. The affine group $W_a(B_3)$ can be generated by the generators $\langle r_0, r_1, r_2, r_3 \rangle$ where $\langle r_1, r_2, r_3 \rangle$ are given in (30) and $r_0 = t(\tilde{\alpha}) r_{\tilde{\alpha}}$ again with $\tilde{\alpha} = e_1 + e_2$ as in Fig. 6d.

A general vector of the root lattice then will be given by $p = b_1 \alpha_1 + b_2 \alpha_2 + b_3 \alpha_3, b_i \in \mathbf{Z}$. This indicates that a general vector of the lattice is the linear combinations of the imaginary quaternions $e_i$ with integer coefficients. It can also be represented as linear combinations of weight vectors $q = a_1 \omega_1 + a_2 \omega_2 + a_3(2\omega_3), a_i \in \mathbf{Z}$. The primitive cell of the lattice can be chosen as the cube with the vertices

$$0, e_i, e_i + e_j \, (i<j), e_i + e_j + e_k \, (i<j<k). \tag{34}$$

There are $2^3$ such cubes sharing the origin as a vertex. Denote by $V(0)$ the Voronoi cell around the origin. The vertices of the Voronoi polyhedron $V(0)$ can be determined as the intersection of the planes surrounding the origin. These planes are determined as the orbits of the fundamental weights

$$\frac{\omega_1}{2} = \frac{e_1}{2}, \frac{\omega_2}{2} = \frac{e_1 + e_2}{2}, \omega_3 = \frac{1}{2}(e_1 + e_2 + e_3). \tag{35}$$

The Voronoi polyhedron $V(0)$ ( the Wigner- Seitz cell) is then the cube around the origin with the vertices

$$(001)_{B_3} = \frac{1}{2}(\pm e_1 \pm e_2 \pm e_3). \tag{36}$$

**4. Constructions of the pseudoicosahedron and its dual polyhedron from $D_3$ with the pyritohedral symmetry**

Two dual platonic solids, the icosahedron and the dodecahedron possess the 5-fold symmetry in addition to 3-fold and 2-fold symmetries. The full symmetry of these polyhedra is the icosahedral symmetry of order 120 which can be derived from the noncrystallographic Coxeter diagram $H_3$. Its quaternionic description follows the same procedure studied in the previous sections. Here one introduces the binary icosahedral subgroup of quaternions by $I = T \cup S$ with

$$I = \langle p, q \rangle, \; p = \frac{1}{2}(e_1 + \tau e_2 + \sigma e_3), \; q = \frac{1}{2}(1 + e_1 + e_2 + e_3) \quad \text{and} \quad \tau = \frac{1+\sqrt{5}}{2} \text{ and } \sigma = \frac{1-\sqrt{5}}{2}$$

(Koca et al, 2007) where T is given in (5) and the set S represents 96 remaining quaternions of the set I. The Coxeter group can be expressed as $W(H_3) = [I, \bar{I}] \cup [I, -\bar{I}]$. The subgroup $[I, \bar{I}] \approx A_5$ is isomorphic to the group generated by even permutations of five letters. It is clear that the Coxeter group is nothing other than the icosahedral group $W(H_3) = A_5 \times C_2$ involving rotations as well as inversions. The set of vertices of the icosahedron is simply the orbit represented by the symbol $(001)_{H_3}$. Since icosahedral symmetry involves the 5-fold symmetry it is not compatible with the crystallography which involves only 2, 3, 4 and 6- fold symmetries only. However, there have been many discoveries in the quasicrystallography displaying the icosahedral symmetry (Shechtman et al., 1984). It was also shown that the quasicrystal with icosahedral symmetry can be obtained by orthogonal projection of the 6-dimensional lattices (Koca et al., 2015). Our discussions here will not follow the quasicrystal structures but 3D crystals with the pyritohedral symmetry which can be derived from the Coxeter-Dynkin diagram $D_3$. Here the pseudoicosahedron plays an essential role in understanding the crystallographic structures with the pyritohedral symmetry. We first note the fact that the pyritohedral symmetry $T_h = \{[T, \bar{T}] \cup [T, -\bar{T}]\}$ consists of only 3-fold and 2-fold symmetries and it is a maximal subgroup of both the octahedral group $W(B_3)$ and the icosahedral group $W(H_3)$. We will prove that by using the pyritohedral symmetry it is possible to derive the vertices of pseudoicosahedron as well as the icosahedron from the diagram $D_3$.

### 4.1. Derivation of the vertices of the pseudoicosahedron from $D_3$ diagram

The rotation generators $r_1 r_2$ and $r_1 r_3$ of $D_3$ generate the subgroup $[T, \bar{T}] \approx A_4$ of order 12. This is the rotational symmetry of a cube with stripes on its faces as shown in Fig. 11. The chiral tetrahedral group $[T, \bar{T}] \approx A_4$ involves 8 rotations by $120^0$ around the 4 diagonals of the cube, 3 rotations by $180^0$ around the $x, y$ and $z$ axes and the unit element. Pyrite crystals often occur in the shape of cubes with striated faces, octahedra and pyritohedra ( a solid similar to dodecahedron but with non regular pentagonal faces), the shape of which will be discussed in this section.

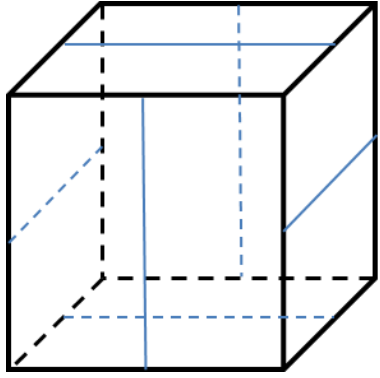

**Figure 11**. The cube with stripes on its faces.

Imposing the Dynkin diagram symmetry $\gamma$ the group is extended to the pyritohedral symmetry $T_h = A_4 \times C_2 = \langle r_1 r_2, r_1 r_3, \gamma \rangle$. Denote by a general vector of $D_3$ by $\lambda = a_1 \omega_1 + a_2 \omega_2 + a_3 \omega_3$ where $\omega_1, \omega_2$ and $\omega_3$ are the weight vectors given in equation (21). We note that the orbit of the pyritohedral group generated by the vector $\lambda$ with $a_1 = 1$, $a_2 = a_3 = 0$ is an octahedron. The orbit generated from $\lambda = \omega_2$ or $\lambda = \omega_3$ corresponds to a cube. Similarly the vector $\lambda = \omega_2 + \omega_3$ leads to the cuboctahedron under the pyritohedral symmetry.

From a general vector we can generate two equilateral triangles sharing the vertex $\lambda$. The vertices $\lambda$, $r_1 r_2 \lambda$ and $(r_1 r_2)^2 \lambda$ form an equilateral triangle with an edge length squared $a_1^2 + a_1 a_2 + a_2^2$. Similarly, the vertices $\lambda$, $r_1 r_3 \lambda$ and $(r_1 r_3)^2 \lambda$ form the second triangle with the edge length squared is $a_1^2 + a_1 a_3 + a_3^2$. If one draws a line between the vertices $\lambda$ and $r_2 r_3 \lambda$ its length squared will be $a_2^2 + a_3^2$. These vectors constitute five vertices $r_1 r_2 \lambda$, $(r_1 r_2)^2 \lambda$, $r_1 r_3 \lambda$, $(r_1 r_3)^2 \lambda$ and $r_2 r_3 \lambda$ surrounding the vertex $\lambda$ as shown in Fig. 12. If we impose all the edge lengths are the same then one obtains five equilateral triangles around one vertex. This would lead to the set of equations

$$a_1^2 + a_1 a_2 + a_2^2 = a_1^2 + a_1 a_3 + a_3^2 = a_2^2 + a_3^2. \tag{37}$$

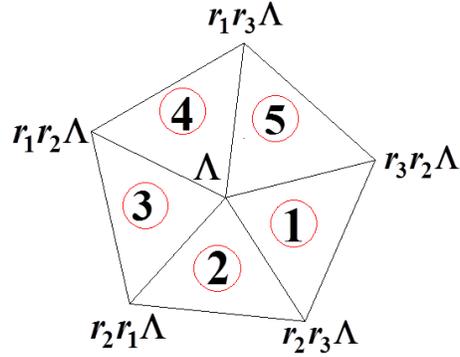

**Figure 12.** Five triangles meeting at one vertex.

Let us assume that the general vector $\lambda$ is invariant under the Dynkin diagram symmetry, that is, $\gamma\lambda = \lambda$. This would imply that $a_2 = a_3$ and a general vector can be written as

$$\lambda = a_1(\omega_1 + x(\omega_2 + \omega_3)) = a_1((1+x)e_1 + xe_2) \qquad (38)$$

where $x = \dfrac{a_2}{a_1} = \dfrac{a_3}{a_1}$ is the parameter which could be computed from (37). Factoring (37) by $a_1$ one obtains the equation $x^2 - x - 1 = 0$. The solution of this equation is either $\tau = \dfrac{1+\sqrt{5}}{2}$ or $\sigma = \dfrac{1-\sqrt{5}}{2}$. The action of the pyritohedral group on the vector given in (38) apart from the scale factor $a_1$ will generate the set of 12 vectors:

$$\pm(1+x)e_1 \pm xe_2, \pm(1+x)e_2 \pm xe_3, \pm(1+x)e_3 \pm xe_1. \qquad (39)$$

Substituting $\tau$ and $\sigma$ respectively for $x$ we obtain two sets of 12 vertices as:

$$\tau\{\pm\tau e_1 \pm e_2, \pm\tau e_2 \pm e_3, \pm\tau e_3 \pm e_1\}, \qquad (40a)$$

$$\sigma\{\pm\sigma e_1 \pm e_2, \pm\sigma e_2 \pm e_3, \pm\sigma e_3 \pm e_1\}. \qquad (40b)$$

These sets of vertices represent two mirror images of an icosahedron albeit a scale factor difference. Multiplying the vertices in (40b) by $\tau^3$ one obtains the following set of quaternions

$$\tau\{\pm e_1 \pm \tau e_2, \pm e_2 \pm \tau e_3, \pm e_3 \pm \tau e_1\}. \qquad (41)$$

The icosahedron represented by (40a) is depicted in Fig. 13.

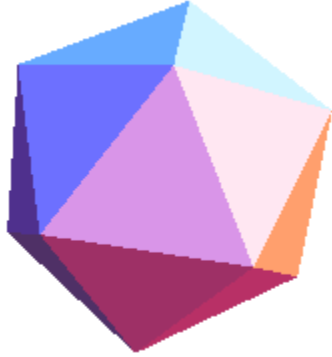

**Figure 13.** The icosahedron.

It can be easily understood that the set of vertices in (41) can be obtained from (40a) by applying any one of the generator of $W(D_3)$, say $r_1$, which interchanges $e_1 \leftrightarrow e_2$ but fixes $e_3$. We should emphasize that the icosahedron is not a chiral solid because (40a) can be transformed to (41) by an orthogonal transformation. Take, for example, an element $[\frac{1}{\sqrt{2}}(1+e_1), \frac{1}{\sqrt{2}}(1-e_1)] \in [T', \bar{T}']$ from the chiral octahedral group. This fixes $e_1$ but interchanges $e_2 \leftrightarrow e_3$ leading to a transformation between two sets of vertices of (40a) and (41).

We have shown in the reference (Koca et al, 2011) that the vertices of the dual of the icosahedron, say, the set of vectors of (40a) can be determined as

$$\frac{1}{2}\{\pm \sigma e_1 \pm \tau e_2, \pm \sigma e_2 \pm \tau e_3, \pm \sigma e_3 \pm \tau e_1\}, \tag{42a}$$

$$\frac{1}{2}(\pm e_1 \pm e_2 \pm e_3). \tag{42b}$$

The 20 vertices in (42a-b) represent a dodecahedron as shown in Fig. 14. Its mirror image can be obtained by replacing $\sigma \leftrightarrow \tau$.

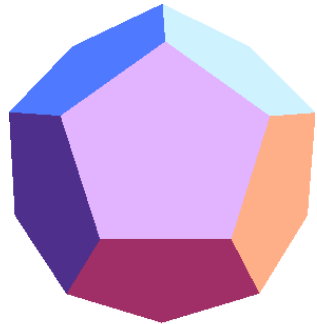

**Figure 14.** The dodecahedron, dual of the icosahedron.

The vectors in (42b) represent the vertices of a cube which is invariant under the pyritohedral symmetry. Similarly the other 12 vertices of (42a) form another orbit under the pyritohedral symmetry. One may now wonders what solid do they represent? Actually they can be obtained from (39) by substituting $x = -\tau$. We note that when $x = -\tau$ the equation in (37) is violated and we obtain two different edge lengths. This follows from the relation $a_1^2 + a_1 a_2 + a_2^2 = a_2^2 + a_2 a_3 + a_3^2 \neq a_2^2 + a_3^2$ leading to two classes of triangles surrounding the vertex $\lambda$: two are equilateral triangles the other three are isosceles triangles. The length squared of the edges of the equilateral triangles apart from a factor $a_1^2$ are $1 + x + x^2$ and the length squared of the isosceles triangles are $1 + x + x^2$, $1 + x + x^2$ and $2x^2$. Therefore for $x = -\tau$ the triangles around the vertex $\lambda$ are either equilateral triangles with edge of 1 unit or isosceles triangles with sides $1, 1, \tau$ known as Robinson triangles. For $x = -\sigma$ the isosceles triangles have the edge lengths proportional to the numbers $1, 1, -\sigma$. It is interesting to observe that the vertices of a dodecahedron split into two orbits under the pyritohedral group, representing a cube and a pseudoicosahedron. Let us define the function $f(x) = x^2 - x - 1$. A plot of this function is shown in Fig.15.

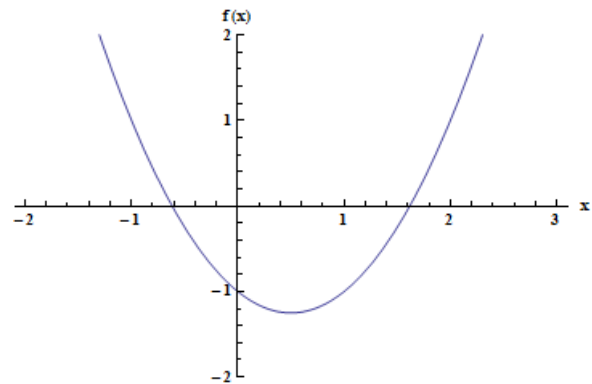

**Figure 15.** The plot of the function $f(x) = x^2 - x - 1$.

We note that for $\sigma \langle x \langle \tau$ where $x \neq 0$ and $x \neq -\dfrac{1}{2}$ the edge with length squared $2x^2$ of the isosceles triangle is less than the edge with length squared $1 + x + x^2$ corresponding to an angle less than $60^0$ subtending the short edge. For $x \langle \sigma$ and $x \rangle \tau$ with $x \neq -1$ we have the inequality $2x^2 \rangle (1 + x + x^2)$ and the angle at the symmetry vertex of the isosceles triangle is greater than $60^0$.

### 4.2. Derivation of the vertices of the pyritohedron from the pseudoicosahedron

We can compute the dual of the pseudoicosahedron of (39) for an arbitrary value of $x$ .(Since the dual of the icosahedron is the dodecahedron the dual of the pseudoicosahedron could be called pseudododecahedron instead of pyritohedron!). This can be achieved by determining the vectors normal to the faces of the pseudoicosahedron of (39). Among the 20 faces of the pseudoicosahedron 8 faces are equilateral triangles and 12 faces are isosceles triangles. From Fig. 12 the center of the triangle with the vertices $\lambda$, $r_1 r_2 \lambda$, $(r_1 r_2)^2 \lambda$ can be taken as $\omega_3$ because $r_1 r_2$ is a rotation around $\omega_3$ for $r_1 r_2 \omega_3 = \omega_3$. Similarly the center of the other equilateral triangle can be taken as $\omega_2$. As we have pointed out before, the vertices generated by the pyritohedral group from either $\omega_2$ or $\omega_3$ would lead to the vertices of a cube given in (42b). The vectors normal to the isosceles triangles can be computed as:

$$b_1 = e_1 + (1+x)e_2,$$
$$b_4 = (1+x)e_1 + e_3, \qquad (43)$$
$$b_5 = (1+x)e_1 - e_3.$$

Note that the relations $r_2 r_1 b_1 = r_3 r_2 b_5 = b_4$, $\gamma b_1 = b_1$ and $\gamma b_4 = b_5$ imply that the vectors $b_1, b_4$ and $b_5$ are in the same orbit under the pyritohedral group. They form a plane orthogonal to the vector $d = (1+x)e_1 + xe_2$ to which $\omega_3 - \omega_2 = e_3$ is also orthogonal. If these five vertices were to determine the same plane then one can show that $((\rho b_1 - \omega_2), d) = 0$ is necessary and sufficient condition. From this relation one obtains $\rho = \dfrac{1+2x}{2(1+x)^2}$. Then the vectors $\rho b_1, \rho b_4, \rho b_5, \omega_2$ and $\omega_3$ determine a pentagon, non-regular in general. Applying the pyritohedral group on these vertices one obtains the set of vectors in two orbits, one set with 12 vectors and the other with 8 vectors:

$$\dfrac{(1+2x)}{2(1+x)^2}\{\pm e_1 \pm (1+x)e_2, \pm e_2 \pm (1+x)e_3, \pm e_3 \pm (1+x)e_1\},$$
$$\dfrac{1}{2}(\pm e_1 \pm e_2 \pm e_3). \qquad (44)$$

These are the vertices of the dual solid (pseudododecahedron) of the pseudoicosahedron of (39). It is generally called pyritohedron which is made of 12 pentagonal faces. A simpler form of this can be obtained by defining the parameter $h = \dfrac{x}{x+1}$. Dropping an overall factor $\dfrac{1}{2}$ then (44) will read

$$\{\pm(1-h^2)e_1 \pm (1+h)e_2, \pm(1-h^2)e_2 \pm (1+h)e_3, \pm(1-h^2)e_3 \pm (1+h)e_1\},$$
$$(\pm e_1 \pm e_2 \pm e_3). \qquad (45)$$

Actually the vertices are to be multiplied by $\dfrac{a_1}{2}$. The vertices of the pyritohedron are usually given in the form of (45). There are several cases that do not follow the line of pseudoicosahedron-pyritohedron duality. For example the vertices in (39) for $x = 0$ represent an octahedron, however, (44) represents the union of two orbits corresponding to a cuboctahedron and a cube, not dual to an octahedron. As shown in Fig. 16 in the limit $x \to 0$ the pseudoicosahedron transforms to an octahedron.

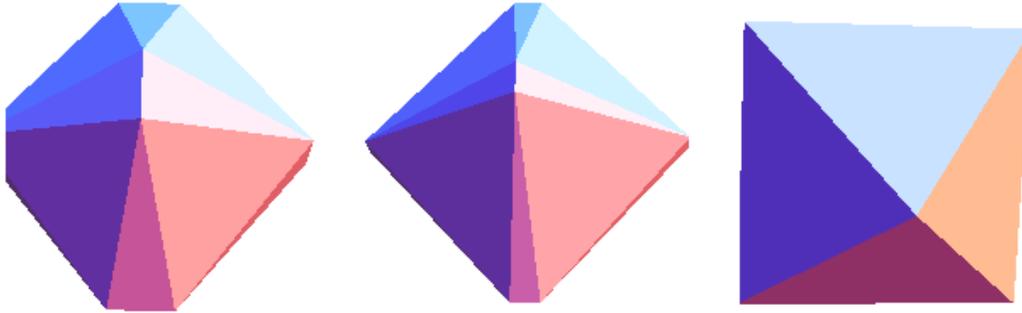

**Figure 16.** Pseudoicosahedron looks like an octahedron in the limit $x \to 0$.

Another interesting limit is that as $x \to \infty$ the vertices of the pseudoicosahedron in (39) transforms to a cuboctahedron. In the same limit $h \to 1$ the vertices in (45) represent a rhombic dodecahedron as shown in Fig. 8.
As for $x = -\tau$ and $x = -\sigma$ the set of vertices of (39) representing two mirror images of a pseudoicosahedon and their duals are obtained from (45) by substituting corresponding values as $h = \tau^2$ and $h = \sigma^2$ respectively representing two mirror images of a pyritohedron.

### 4.3. Pseudoicosahedron and pyritohedron for lattices

Now we would like to discuss pseudoicosahedra and its dual pyritohedra relevant to the crystallography. If we choose $x = 1$ corresponding to $h = \dfrac{1}{2}$ the set of vertices of the pseudoicosahedron in (39) would read

$$(\pm 2e_1 \pm e_2, \pm 2e_2 \pm e_3, \pm 2e_3 \pm e_1). \tag{46}$$

This is a pseudoicosahedron with isosceles triangles of edges $1, 1, \sqrt{\dfrac{2}{3}}$ if $a_1 = 1$. When we substitute $h = \dfrac{1}{2}$ in (45) we obtain the vertices of the pyritohedron dual to the pseudoicosahedron in (46) up to a scale factor.
$$\{\pm 3e_1 \pm 6e_2, \pm 3e_2 \pm 6e_3, \pm 3e_3 \pm 6e_1\}, 4(\pm e_1 \pm e_2 \pm e_3). \tag{47}$$

The set of vertices in (46) and (47) belong to the simple cubic lattice. They are shown in Fig. 17.

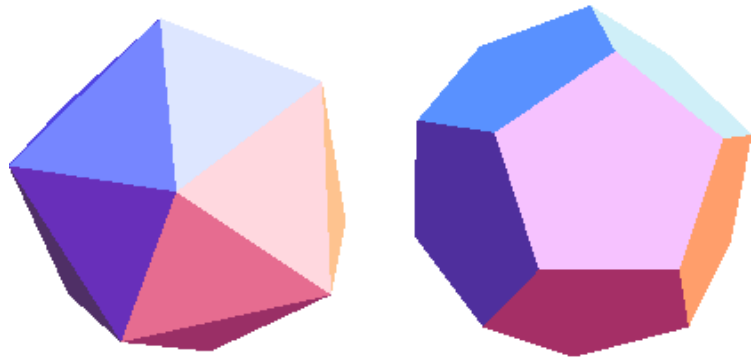

**Figure 17**.  The pseudoicosahedron for $x=1$ and its dual pyritohedron.

Many more candidates of pseudoicosahedra and its dual pyritohedra can be obtained as a structure in the simple cubic lattice. The unit cubic cell can be stacked in the pseudoicosahedron or in the pyritohedron as long as the vertices are chosen from the simple cubic lattice. One wonders whether the mineral (Özgür, N., 1993) found in Fig. 18 has this type of isosceles triangles.

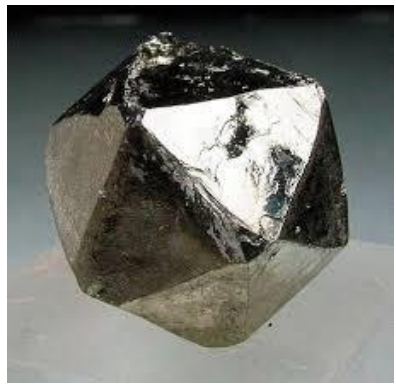

**Figure 18.** Pseudoicosahedral pyrite (Cakmakkaya Mine, Murgul Cu-Zn-Pb deposit, Murgul, Artvin Province, Black Sea Region, Turkey).

The union of the pseudoicosahedron in (46) and its mirror image form the vertices of a truncated octahedron which is the Wigner-Seitz cell of the bcc lattice:

$$\pm 2e_1 \pm e_2, \pm 2e_2 \pm e_3, \pm 2e_3 \pm e_1,$$
$$\pm e_1 \pm 2e_2, \pm e_2 \pm 2e_3, \pm e_3 \pm 2e_1. \tag{48}$$

Let us recall that the simple cubic lattice is a sublattice of the bcc lattice. The truncated octahedron with the vertices of (48) is shown in Fig. 9

For a general $x$ the union of two pseudoicosahedra

$$\pm(1+x)e_1 \pm xe_2, \pm(1+x)e_2 \pm xe_3, \pm(1+x)e_3 \pm xe_1,$$
$$\pm xe_1 \pm (1+x)e_2, \pm xe_2 \pm (1+x)e_3, \pm xe_3 \pm (1+x)e_1 \quad (49)$$

represents the vertices of a non-regular truncated octahedron which can be derived as the orbit of the Coxeter group $W(B_3)$ denoted by $a_1(1,x,0)_{B_3}$. This belongs to the simple cubic lattice if $a_1$ and $a_1 x = a_2$ are integers which implies that $x$ should be a rational number. In other words as long as $a_1$ and $a_2$ are integers the pseudoicosahedron with the vertices

$$\pm(a_1+a_2)e_1 \pm a_2 e_2, \pm(a_1+a_2)e_2 \pm a_2 e_3, \pm(a_1+a_2)e_3 \pm a_2 e_1 \quad (50)$$

can be embedded in a simple cubic lattice. Of course its mirror image obtained by interchanging $e_1 \leftrightarrow e_2$ in (50) also represents a pseudoicosahedron embedded in the simple cubic lattice.

### 4.4. Fibonacci sequence of pseudoicosahedra

Fibonacci sequence represents the numbers in the following integer sequence: 1,1,2,3,5,8,13,21,…which can be represented by the recurrence relation $F_n = F_{n-1} + F_{n-2}$ with values $F_1 = F_2 = 1$. The general term of the Fibonacci sequence can also be written as

$$F_n = \frac{\tau^n - \sigma^n}{\tau - \sigma} \quad . \quad (51)$$

The ratio of Fibonacci sequence numbers converges to the golden ratio $\tau$:

$$\lim_{n \to \infty} \frac{F_{n+1}}{F_n} = \tau \quad (52)$$

Let us consider the interval $1 \langle x \langle 2$ in Fig. 15 and define the sequence that the parameter $x$ takes the rational numbers $x_n = \frac{F_{n+1}}{F_n}$ such as

$$1, \frac{2}{1}, \frac{3}{2}, \frac{5}{3}, \frac{8}{5}, \frac{13}{8}, \frac{21}{13}, \frac{34}{21}, \frac{55}{34}, \frac{89}{55}, \ldots \quad (53)$$

The sequence of (53) takes values in the interval $1 \langle x \langle 2$ and approaches to the golden ratio $\tau$. This leads to an infinite sequence of pseudoicosahedra in the simple cubic lattice with the appropriate choice of the integer parameter $a_1$. For example, choosing $a_1 = 3$ for $x_4 = \frac{5}{3}$ then the vertices of the pseudoicosahedron would read:

$\pm 8e_1 \pm 5e_2, \pm 8e_2 \pm 5e_3, \pm 8e_3 \pm 5e_1$. For the mirror pseudoicosahedron just change $e_1 \leftrightarrow e_2$. Similar consideration can be made for the pyritohedron. We list some of those pseudoicosahedra following the Fibonacci sequence by showing only components of the vectors:

$$x_1 = \frac{1}{1}; \ \{(\pm 2, \pm 1, 0), (0, \pm 2, \pm 1), (\pm 1, 0, \pm 2)\},$$

$$x_2 = \frac{2}{1}; \ \{(\pm 3, \pm 2, 0), (0, \pm 3, \pm 2), (\pm 2, 0, \pm 3)\},$$

$$x_3 = \frac{3}{2}; \ \{(\pm 5, \pm 3, 0), (0, \pm 5, \pm 3), (\pm 3, 0, \pm 5)\},$$

$$x_4 = \frac{5}{3}; \ \{(\pm 8, \pm 5, 0), (0, \pm 8, \pm 5), (\pm 5, 0, \pm 8)\}.$$

(54)

It is also interesting to note that the centers of the edges of the pseudoicosahedron of (39) represent 30 vertices of a pseudoicosidodecahedron given in the form of two orbits under the pytitohedral group by:

$$a_1\{\pm xe_1 \pm (1+x)e_2 \pm (1+2x)e_3, \pm (1+x)e_1 \pm (1+2x)e_2 \pm xe_3, \pm (1+2x)e_1 \pm xe_2 \pm (1+x)e_3\},$$

$$a_1(1+x)\{\pm e_1, \pm e_2, \pm e_3\}.$$

(55)

In the limit $x \to \tau$ the vertices in (55) represent a regular icosidodecahedron with two regular pentagons and two equilateral triangles meeting at every vertex as shown in Fig. 19(a). Otherwise the regular pentagons are replaced by irregular pentagons (Fig. 19(b)).

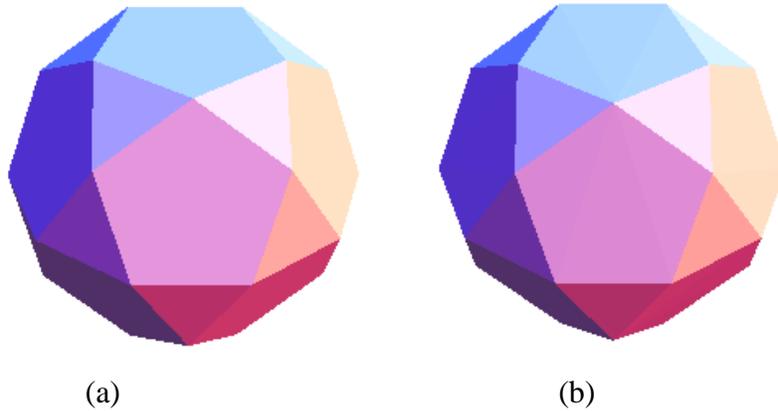

(a)          (b)

**Figure 19**.(a) An icosidodecahedron with regular pentagons and equilateral triangular faces, (b) A pseudoicosidodecahedron obtained from(55) for $x = \dfrac{3}{2}$.

## 5. Discussion

This work is an attempt to construct the cubic lattices and their symmetries with quaternions. We have identified quaternionic group elements with the symmetry axis of a cube including inversion with respect to the origin. Not only the group elements but also the lattice vectors are represented by quaternions. Since the quaternion group is isomorphic to the unitary group $SU(2)$ normally one can describe the spin states of an electron and its symmetry group by quaternions. This gives us a framework to discuss two different physical phenomena, lattices and spin $\dfrac{1}{2}$ states with the same mathematical structure.

In addition to the quaternionic representation of the lattice structures and their symmetries, in this work, we worked out in detail the pyritohedral crystallographic lattice structures. The polyhedra possessing the pyritohedral symmetry have been constructed in terms of quaternions and pointed out that the Fibonacci sequences of the pseudoicosahedron, its dual pyritohedron and pseudoicosidodecahedron approach to their limiting regular polyhedra possessing five-fold symmetry in addition to the pyritohedral symmetry. Although the crystals with pyritohedral symmetry exist in the form of stratified cube, octahedron and pyritohedron, one wonders, why one does not see many other minerals in the form of pseudoicosahedron except a single one depicted in Fig. 18. It is natural to expect that the crystal structures in the form of pseudoicosahedron as well as pseudoicosidodecahedron may exist.